\documentclass[12pt,preprint]{aastex}

\newcommand{\teff}{T$_{\rm eff}$}

\begin{document}

\title{Magnesium isotope ratios in Hyades 
stars\footnote{Data presented herein were obtained at the W.~M.~Keck Observatory, which is 
operated as a scientific partnership among the California Institute of Technology, the 
University of California, and the National Aeronautics and Space Administration. The 
Observatory was made possible by the generous financial support of the W.M. Keck Foundation.}
}

\author{David Yong, David L. Lambert, Carlos Allende Prieto}
\affil{Department of Astronomy, University of Texas, Austin, TX 78712}
\email{tofu,dll,callende@astro.as.utexas.edu}

\and

\author{Diane B. Paulson}

\affil{Department of Astronomy, University of Michigan, 830 Dennison Building, Ann Arbor, MI 48109}
\email{apodis@umich.edu}

\begin{abstract}
Using classical model atmospheres and an LTE analysis, Mg isotope ratios 
$^{24}$Mg:$^{25}$Mg:$^{26}$Mg are measured in 32 Hyades dwarfs covering
4000K $\le$ \teff~$\le$ 5000K.  We find no significant trend in any isotope ratio versus
\teff~and the mean isotope ratio is in excellent agreement with the solar value.  We determine
stellar parameters and Fe abundances for 56 Hyades dwarfs covering 4000K $\le$ \teff~$\le$ 6200K.
For stars warmer
than 4700K, we derive a cluster mean value of [Fe/H] = $0.16\pm 0.02~(\sigma=0.1)$, in good
agreement with previous studies.  
For stars cooler than 4700K, we find that the abundance of Fe from ionized lines exceeds
the abundance of Fe from neutral lines.  At 4700K [Fe/H]$_{\rm II}-$[Fe/H]$_{\rm I}\simeq$ 0.3
dex while at 4000K [Fe/H]$_{\rm II}-$[Fe/H]$_{\rm I}\simeq$ 1.2 dex.
This discrepancy between the Fe abundance from neutral and ionized lines
likely reflects inadequacies in the model atmospheres and the presence of Non-LTE or other 
effects.  Despite
the inability of the models to reproduce ionization equilibrium for Fe, the Mg isotope ratios
appear immune to these problems and remain a powerful tool for studying Galactic chemical 
evolution.  

\end{abstract}


\section{Introduction}

Open clusters provide an ideal opportunity to test our understanding of stellar structure and 
evolution.  Within a given cluster, the individual stars may span a considerable range in mass 
and evolutionary state.  Cluster stars 
are believed to have formed at the same time from a chemically homogeneous reservoir of gas.  
The Hyades open cluster has been the focus of many studies due to its proximity and the importance
of the Hyades extends beyond stellar structure and evolution.  The seminal papers by 
\citet{perryman98} and \citet{debruijne01} address key issues including membership, distance, and 
age as well as outlining the prominent role played by the Hyades over the past century.

While it had been assumed that all stars within a cluster have the same composition, 
\citet{conti65} were the first to show that cluster stars are chemically homogeneous (with
Li \citep{boesgaard86} and Be \citep{boesgaard02} being notable exceptions) 
through an analysis of 10 Hyades dwarfs.  As a result of the first
dredge up, cluster subgiants and giants will have different compositions than unevolved stars.  
\citet{varenne99} review the subsequent abundance determinations
for Hyades stars.  Recently, \citet{paulson03} have confirmed the uniformity of the iron abundance
through an analysis of 55 Hyades dwarfs with spectral types ranging from F to K.  
\citeauthor{paulson03} have also shown that within the measurement uncertainties, the abundance 
ratios of Na, Mg, Si, Ca, Ti, and Zn with respect to Fe are constant (at the 1$\sigma$ = 0.04 dex 
level) and in their solar proportions.

In this paper, we utilize the chemical homogeneity of Hyades stars to study 2 related issues.  
The primary question we seek to answer is whether or not Hyades stars have uniform Mg isotopic 
ratios?  In \citet{mghdwarf}, we measured the Mg isotope ratios in cool field dwarfs to study Galactic 
chemical evolution.  There was a scatter in the isotope ratios at a fixed [Fe/H] and a hint
of an increasing ratio with decreasing \teff.  It was also curious that the 2 stars with remarkably 
high ratios of $^{25,26}$Mg/$^{24}$Mg had particularly strong MgH lines.  Our assumption is that 
the true Mg isotope ratios are uniform in Hyades dwarfs.  An observational contradiction of this
assumption is here taken as a failure of the analytical technique.
The second concern to be addressed is how appropriate are the model atmospheres and 
the assumptions underpinning the analysis when studying cool stars?  The analysis of a chemically 
homogeneous sample of stars spanning a broad range in effective temperature 
(\teff) is a powerful test of a model atmosphere grid and the analysis techniques.  Departures from local 
thermodynamic equilibrium (LTE) may manifest as \teff-dependent 
abundance anomalies.  In particular, we note that 
\citet{feltzing98}, \citet{schuler03} and Allende Prieto et.\ al (2003, 
private communication) have studied cool stars and found an overionization of Fe with respect to
the LTE predictions.  The degree of 
overionization increases with decreasing \teff.  All these studies rely upon classical one-dimensional 
LTE model atmospheres.  Recent advances in model atmospheres include non-local thermodynamic 
equilibrium (Non-LTE) models (e.g.\ \citealt{nextgen99}) and LTE three-dimensional time-dependent 
hydrodynamical models (e.g.\ \citealt{asplund00}).  Unfortunately, construction of these new models 
requires large amounts of computing time and, therefore, such models only cover a small range of 
stellar parameters.  

Abundance analyses of cool stars are rare, and 
no previous study of the Hyades has investigated stars cooler than \teff$\simeq$4800K.  
Hyades stars are perfect targets for investigating any dependence of the Mg isotope ratios upon
\teff~as well as identifying whether the problem of overionization of Fe exists in cool Hyades dwarfs
when using classical model atmospheres.  
Here we measure the Mg isotope ratios in 32 Hyades dwarfs with 4000K $\le$ \teff~$\le$ 5000K.  (For
stars warmer than 5000K, the MgH lines -- from which the Mg isotope ratios are measured -- are too weak.)
We also determine stellar parameters and iron abundances for 56 Hyades dwarfs with 4000K 
$\le$ \teff~$\le$ 6200K.  This is the first study of the Mg isotope ratios in an open cluster and
incorporates the coolest Hyades stars to which detailed spectroscopic abundance analyses have been 
applied.

\section{Observations and data reduction}

The stars in this paper are a subset of those being analyzed as part of a planet search program.  
For the complete description of observations and candidate selection, see \citet{cochran02}.  The 
observations were made using HIRES \citep{hires} on the Keck I telescope between 1996 and 2002 with 
a resolving power of $R \equiv \lambda/\Delta\lambda = 60,000$.  The wavelength range (3800 to 
6200\AA) was selected to include I$_2$ absorption lines and to monitor stellar chromospheric activity 
in Ca\,{\sc ii} H and K lines as required by the radial velocity 
program.  The spectra used in this study are the ``template''
spectra which were taken without the I$_2$ cell in the stellar beam and therefore are 
free of I$_2$ absorption lines.  The signal-to-noise ratio (S/N) ranged from 80 to
200 per pixel at 5140\AA.  The MgH lines from which the Mg isotope ratios will be derived are
located at 5140\AA.  One dimensional wavelength calibrated normalized spectra were extracted 
in the standard way using the IRAF\footnote{IRAF is distributed by the National Optical Astronomy 
Observatories, which are operated by the Association of Universities for Research in Astronomy, 
Inc., under cooperative agreement with the National Science Foundation.} package of programs.  

\section{Stellar parameters and the iron abundance}

\teff~were determined using the \citet{alonso96b} \teff:[Fe/H]:color relations based on the 
infrared flux method.  We used the Str{\" o}mgren $b-y$ index and the $B-V$ index where 
\citeauthor{alonso96b} state that the standard deviation from the fits are 110K and 130K 
respectively.  We have $B-V$ \citep{ap99} for all stars and $b-y$ \citep{hauck98} for 
32 of the 56 stars.  We applied the \citeauthor{alonso96b} \teff:[Fe/H]:color relations to 
determine \teff~using both 
color indices assuming a metallicity [Fe/H]=0.10.  If we adopt the \citet{paulson03} metallicity
[Fe/H]=0.13, the values for \teff~would change by 5K.  For the 32 stars with $B-V$ and $b-y$ 
photometry, we adopted the mean \teff.  For these 32 stars, we found that $\langle$\teff$_{b-y} - 
$\teff$_{B-V}\rangle = 24$K ($\sigma=60$K).  For the remaining stars with only $B-V$ photometry, 
we adopted \teff$_{B-V} + 10$K.

In order to determine the surface gravity, we used the 600 Myr solar metallicity $Y^2$ isochrones 
calculated by
\citet{yi01}.  (\citet{perryman98} have shown that the Hyades have an age 625$\pm$50 Myr.) 
We fitted a spline function through the log g and \teff~values given in the isochrone and assumed 
the gravity corresponding to the adopted \teff.  If we use the 800 Myr solar metallicity isochrones,
the gravities would only change by 0.01 dex.

\begin{figure}
\epsscale{1.0}
\plotone{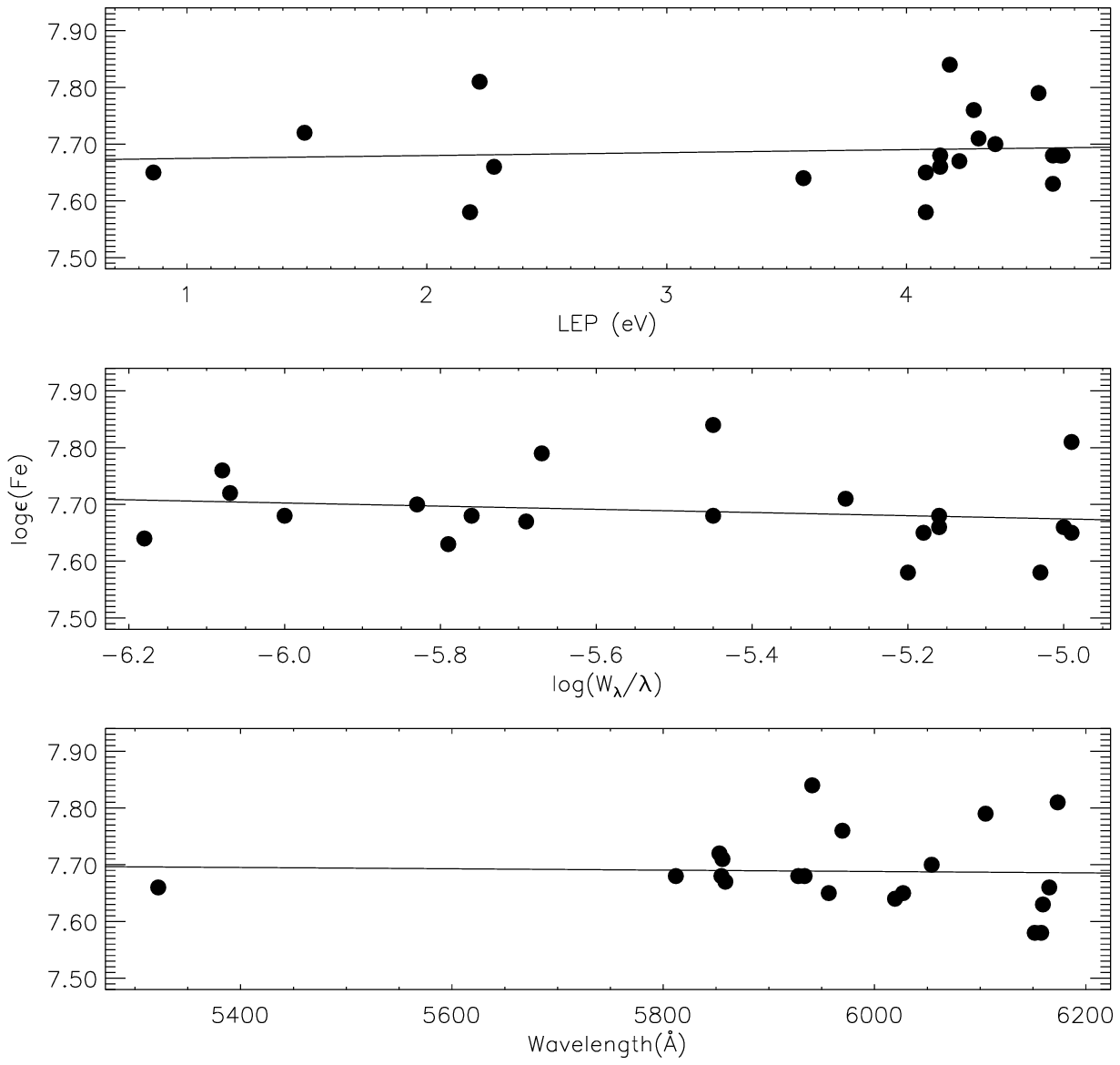}
\caption{Fe abundance from individual Fe\,{\sc i} lines versus excitation potential
(upper), reduced equivalent width (middle), and wavelength (lower).  The lower excitation 
potential (LEP)-abundance relation can be used to set \teff~and the reduced equivalent width 
(W$_\lambda$/$\lambda$)-abundance relation is used to determine $\xi_t$.  In
all panels the line represents the linear least-squares fit to the data.
\label{fig:param.check}}
\end{figure}

Equivalent widths (EWs) were measured for a selection of 20 Fe\,{\sc i} and 9 Fe\,{\sc ii} lines using IRAF
where in general Gaussian profiles were fit to the observed profile.  These lines were identical to those
used by \citet{paulson03} and are presented in Table \ref{tab:lines}.  The $gf$ values were taken from
\citet{kurucz95}, \citet{lambert96}, \citet{schnabel03}, and from a compilation by R.E. Luck 
(1993, private communication).  
The solar Fe abundance was derived from a solar 
spectrum (of Ceres) taken through HIRES.  Due to instrumental effects, there 
was a difference of 0.10 dex between our derived log$\epsilon$(Fe)$_\odot$ and 
that found by \citet{grevesse99}.  For the Hyades stars, we used a line-by-line differential 
analysis to derive [Fe/H].  Such an analysis ensures that errors in the $gf$ values do not 
greatly affect the derived abundances. 
The model
atmospheres were taken from the \citet{kurucz93} LTE stellar atmosphere grid.  We interpolated within
the grid when necessary to produce a model with the required \teff, log g, and [Fe/H].  The model
was used in the LTE stellar line analysis program {\sc Moog} \citep{moog}.  The microturbulence 
($\xi_t$) was determined in the usual way by insisting that the Fe abundance from Fe\,{\sc i} lines 
be independent of their equivalent width ($W_\lambda$).  See Figure \ref{fig:param.check} for an 
example of how the abundance versus $W_\lambda$ plot is used to set $\xi_t$.  While this method worked 
for the warmer stars, we encountered problems with stars cooler than about 4900K where no value of
$\xi_t$ would produce a zero trend in the abundance versus $W_\lambda$ plot.  \citet{feltzing98}
report a similar problem where ``for the K dwarf stars none of the described 
methods seemed to yield definite values for the microturbulence parameter''.  For these stars,
we adopted a microturbulence of 0.3 km s$^{-1}$, a value which reduced (but did not eliminate)
the trend in the abundance versus $W_\lambda$ plot.  Later in the paper we show that our results
are not significantly affected by our choice of microturbulence.  The stellar parameters are given 
in Table \ref{tab:param}.

\begin{figure}
\epsscale{1.0}
\plotone{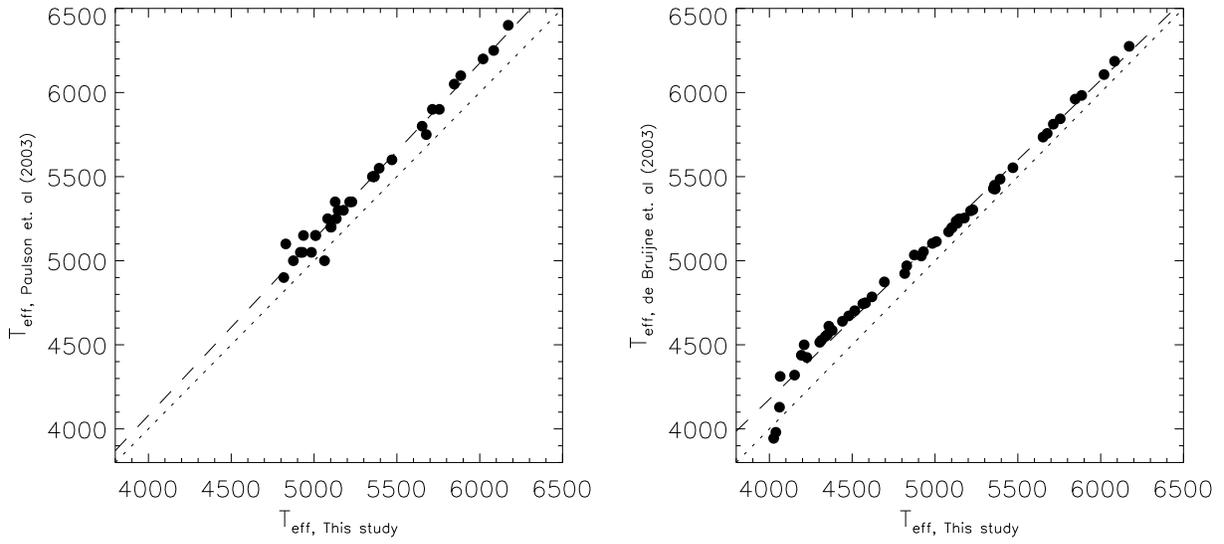}
\caption{\teff~comparison between this study and \citet{paulson03} (left)
and \citet{debruijne01} (right).
The dotted line represents the line of equality while the dashed line
is the linear least-squares fit to the data.
\label{fig:teff.comp}}
\end{figure}

An alternative method for deriving \teff~is by forcing the abundances of individual 
Fe\,{\sc i} lines to be independent of lower excitation potential.  See Figure \ref{fig:param.check} 
for an example of how \teff~can be determined by requiring excitation equilibrium.  While we only
have 5 Fe\,{\sc i} lines with lower excitation potentials less than 3eV, the adopted \teff~generally
result in Fe abundances independent of lower excitation potential.  Many of the stars in this study 
are included in \citet{debruijne01} and \citet{paulson03} and in Figure \ref{fig:teff.comp} and
Table \ref{tab:comp}, we 
compare the values for \teff.  On average, we find that our \teff~are 150K cooler than the 
\citet{paulson03} values and 140K cooler than the \citet{debruijne01} values.  These differences
are comparable to the uncertainties arising from the application of the \citet{alonso96b} 
\teff:[Fe/H]:color relations.  We note that \citeauthor{paulson03} determine \teff~from excitation
equilibrium whereas \citeauthor{debruijne01} derive \teff~(for stars cooler than 7250K) 
from the \citet{lejeune98} calibrations for \teff:$B-V$:log g.  If reddening significantly 
affected the color indices, this could explain why our \teff~are cooler than the \citeauthor{paulson03}
values, but not the \citeauthor{debruijne01} values.  
However, Hyades stars are not significantly reddened with $E(B-V)=0.003 \pm 0.002$ mag \citep{taylor80}.  

\begin{figure}
\epsscale{1.0}
\plotone{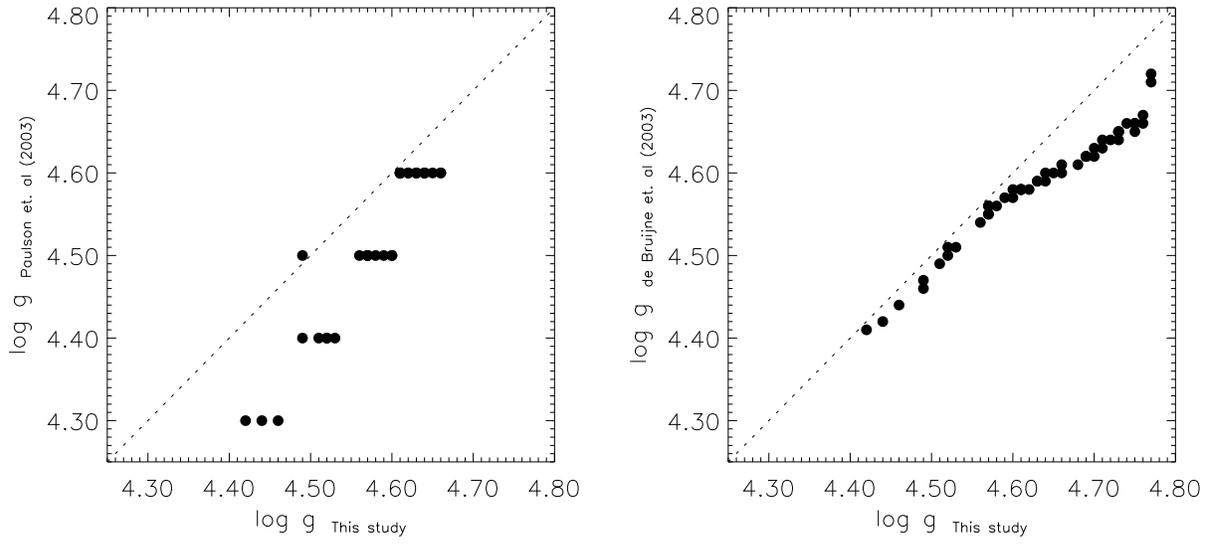}
\caption{Gravity comparison between this study and \citet{paulson03} (left)
and \citet{debruijne01} (right).
The dotted line represents the line of equality.
\label{fig:logg.comp}}
\end{figure}

Surface gravities can be set by requiring the Fe abundance derived from Fe\,{\sc i} lines match the
Fe abundance derived from Fe\,{\sc ii} lines, i.e., ionization equilibrium.  \citeauthor{paulson03}
set their gravities by requiring ionization equilibrium while \citeauthor{debruijne01} use the
\citet{lejeune98} calibrations for \teff:$B-V$:log g.  
In Figure \ref{fig:logg.comp} and Table \ref{tab:comp}, we compare the 
values for surface
gravity.  The agreement is good between the various studies where the maximum difference between this
study and \citeauthor{paulson03} is 0.16 dex while the maximum difference between this study and
\citeauthor{debruijne01} is 0.10 dex.  

A striking result of our analysis is that the Fe abundance from
Fe\,{\sc ii} lines exceeds that from the Fe\,{\sc i} lines.
The excess (Figure 4) is approximately constant for
\teff $> 5000$ K at 0.2 dex, but increases steeply with
decreasing \teff~reaching the remarkable value of
about 1 dex at 4000 K. 
\citet{bdp93} and \citet{bdp03} studied F and G dwarfs 
(5500K $\le$ \teff $\le$ 6500K) and found good, but not perfect, agreement
between the abundances from neutral and ionized iron.  
The abundance from the Fe\,{\sc i}
lines is constant for stars hotter than about 4300 K but
increases by about 0.2 dex for stars of 4000 K. Ionization
equilibrium is not satisfied by our analysis. This result
was anticipated by several earlier analyses but no
previous analysis has examined stars as cool as 4000K.

\citet{paulson03} considered only stars hotter than
4700K.  In this range, imposition of LTE ionization
equilibrium by their spectroscopic method of determining
\teff~and log g does not yield a result for the
Fe abundance significantly different from ours.
Our mean abundance for stars hotter than 4700K is 
[Fe/H]=$0.16\pm 0.02~(\sigma=0.1)$ where all lines were given
equal weight.
In Table \ref{tab:aberr}, we present the abundance 
dependences upon the model parameters.  (Note that our choice of
microturbulence and the adopted surface gravity do not significantly 
affect the results.)
Our abundance difference from neutral and ionized lines
for these warm stars
vanishes with a temperature correction of around 150K
or a change of log g by 0.5 dex.  The \teff~adjustment 
is not implausible, but the gravity reduction
of 0.5 dex may be rejected as inconsistent with
the gravity predicted by the evolutionary tracks. Application
of the temperature correction raises the mean Fe abundance
by about 0.1 dex.  

\citet{schuler03} studied dwarfs in the open cluster M 34   
using high-resolution spectra and Kurucz models
with \teff~set by excitation
equilibrium of Fe\,{\sc i} lines and surface gravities estimated
from an empirical relation. Their stars covered the temperature
range from 4700 to 6200K. At 6200K, the Fe abundance from the
neutral lines is about 0.15 dex higher than that from the
ionized lines, but at 4700K the ionized lines give the higher
abundance by about 0.6 dex. These results are in reasonable
agreement with those in Figure \ref{fig:fe.all}.   

\citet{feltzing98} found overionization ([Fe/H]$_{\rm II}-$[Fe/H]$_{\rm I} \simeq 0.3-0.4$) 
for five K dwarfs in the range 4510K $\le$ \teff~$\le$ 4833K 
with no obvious trend with \teff.  They considered the possibility that the \teff~scale may be in error and 
concluded that departures from LTE are a likely cause of the discrepancy.  \citet{thoren00} reanalyzed the 
five problematic K dwarfs.  Use of a \teff~scale based on excitation equilibrium rather than photometric 
temperatures changed the \teff~of 1 star by +200K.  By fitting the wings of strong lines, surface gravities
were changed in the remaining stars by 0.3 to 0.5 dex.  Synthetic spectra allowed for revision of the 
continuum and a different linelist was employed.  Modification of their analysis techniques showed that the 
overionization of Fe was small, [Fe/H]$_{\rm II}-$[Fe/H]$_{\rm I} \simeq 0.1$.

Figure 4 is evidence that the standard combination of a classical
atmosphere and LTE physics of line formation fails to reproduce
the collection of neutral and ionized iron lines in the spectra
of cool dwarfs with the failure increasing with decreasing
temperature.\footnote{The failure cannot be eliminated by 
alternative sources of classical atmospheres. 
Use of NEXTGEN \citep{nextgen99} models which include improved
molecular opacities does not materially change the results in
Figure \ref{fig:fe.all}.}  We used the van der Waals line damping parameter 
(Uns{\" o}ld approximation multiplied by a factor recommended by the Blackwell group).
We tested other damping parameters and none resulted in a closer agreement between
the abundances from neutral and ionized iron.

A suspicion aired in previous papers is that 
Non-LTE effects on the ionization equilibrium between neutral Fe
atoms and the singly-charged ions are responsible for discrepant
abundances from Fe\,{\sc i} and Fe\,{\sc ii} lines: iron atoms
are over-ionized (relative to LTE) by the ultraviolet radiation
field. Published calculations of Non-LTE effects on iron
atoms and ions (e.g., \citealt{thevenin99,gehren01,shchukina01}) 
for stars of approximately solar
metallicity and of solar or warmer temperatures
indicate the effects are small. Overionization at these temperatures, were it a large
effect, would be seen most obviously as an underabundance in LTE
analyses of the Fe\,{\sc i} with very slight effects on the 
analyses of the Fe\,{\sc ii}; iron is predominantly singly-ionized
in F and G atmospheres. These calculations
use classical (LTE) atmospheres and, hence, are an incomplete
characterization of the Non-LTE effects.  In the
atmospheres of the coolest stars
of our study, neutral atoms greatly outnumber the ions and, hence,
overionization can greatly increase the number of ions without
significantly decreasing the number of neutral atoms.
Overionization is likely driven by the flux of ultraviolet photons
penetrating the line-forming regions. With or even without a
chromosphere, the mean intensity of this flux ($J_\nu$) seems certain to
exceed the flux assumed in LTE (i.e., $B_\nu$, the local Planck
function). The Hyades dwarfs are chromospherically active 
(e.g., \citealt{wilson63,duncan84,rhm95})
and, therefore, likely to provide for greater Non-LTE effects
than in comparable but older stars. 
Quantitative evaluation of the Non-LTE effects on iron and
other elements are awaited with interest.

A possible contributor to the apparent disequilibrium of iron
ionization in the coolest dwarfs is that classical model
atmospheres are an inadequate representation of the real
atmospheres of these dwarfs. Stellar granulation is not
recognized by the classical assumptions. Spots may be prevalent
on these young stars. Heating processes supporting the
temperature rise of the chromosphere may affect the
photospheric structure. Non-LTE effects may influence
the photospheric structure.   

\section{Magnesium isotopic ratios}

Our primary goal is to determine the Mg isotopic ratios in the Hyades
dwarfs from the warmest stars in which MgH lines are of adequate
strength (\teff $\simeq$ 5000K) to the coolest in our
sample (\teff $\simeq$ 4000K). The working assumption is that
the stars are chemically homogeneous, certainly with respect to
the Mg isotopes. Initially, the principal motivation was an
earlier suspicion that measured isotopic ratios from spectra of
dwarfs were dependent on a star's \teff~\citep{mghdwarf}. With 
the discovery of severe ionization disequilibrium
in the coolest Hyades dwarfs, the motivation for pursuing the
Mg isotopic ratios was greatly strengthened.

The isotopic ratios are obtained from MgH lines near 5140\AA. 
Figure \ref{fig:mgh.all} shows a representative spectrum of a star with strong
MgH lines. At a given \teff, stars have identical spectra,
as illustrated in Figure \ref{fig:mgh.comp} where we overplot the spectra of two
stars with almost identical parameters. There are no discernable
differences in the profiles of the MgH features suggesting that
the isotopic ratios may indeed be very similar.  

While many 
MgH lines are present in the spectra of cool stars, few are suitable for isotopic analysis \citep{tl80}.  To 
measure the Mg isotope ratios, we rely upon three MgH features recommended by \citet{ml88} and used by 
\citet{gl2000} and \citet{6752,mghdwarf}.  The three lines are at 5134.6\AA, 5138.7\AA, and 5140.2\AA~and we
refer to them as Region 1, 2, and 3.  These lines are described in detail by \citet{ml88}.  The 
macroturbulence was determined by fitting the profile of an unblended line, Ti\,{\sc i} at 5145.5\AA.  For all
stars, this Ti\,{\sc i} line was slightly stronger than the recommended MgH lines.  The
macroturbulence was assumed to have a Gaussian form representing the approximately equal 
contributions from atmospheric turbulence, stellar rotation,
and instrumental profile.  The linelist was identical to the one used by \citet{gl2000} and includes 
contributions from C, Mg, Sc, Ti, Cr, Fe, Co, Ni, and Y.  The wavelengths of all isotopic components were 
taken from \citet{ml88} and were based on direct measurements of an MgH spectrum obtained using a Fourier 
transform spectrometer by \citet{bernath85}.  

The abundances of $^{25}$Mg and $^{26}$Mg were adjusted until
the profiles of the 3 recommended features were best fit.  Following the work by \citet{nissen99,nissen00}
on Li isotope ratios and \citet{6752} on Mg isotope ratios, we used a $\chi^2$ analysis to determine the
best fit to the data.  The advantages to this method are that it is unbiased and errors in the fits
can be quantified.  The free parameters
were (1) $^{25}$Mg/$^{24}$Mg, (2) $^{26}$Mg/$^{24}$Mg, and (3) log$\epsilon$(Mg) and we treated each of 
the three recommended features independently.  We took the terrestrial ratio, $^{24}$Mg:$^{25}$Mg:$^{26}$Mg =
78.99:10.00:11.01 \citep{mghsolar} as our initial guess and we refer to these ratios as the solar ratios.  
We explored a large range of parameter space around our initial
guess and calculated $\chi^2 = \Sigma(O_i-S_i)^2/\sigma^2$ where $O_i$ is the observed spectrum point, 
$S_i$ is the synthesis, and $\sigma = (S/N)^{-1}$.  The optimum values for $^{25}$Mg/$^{24}$Mg, 
$^{26}$Mg/$^{24}$Mg, and log$\epsilon$(Mg) were determined by locating the minima in $\chi^2$.  
We tested over 500 different synthetic spectra per
region and the minimum $\chi^2_{red}=\chi^2/\nu$, where $\nu$ is the number of degrees of freedom in the 
fit, was sufficiently close to 1.  Examples of spectrum syntheses are shown in Figures \ref{fig:region1} and
\ref{fig:region23} and the optimal isotope ratios are given in Table \ref{tab:param}.  Note that the red 
asymmetry on the MgH lines is due to the presence of $^{25}$MgH and
$^{26}$MgH.  The synthesis computed assuming only $^{24}$MgH provides a poor fit to the spectrum.  

Following \citet{bevington92}, we plotted $\Delta\chi^2 = \chi^2 - \chi^2_{min}$ against the ratios 
$^{25}$Mg/$^{24}$Mg and $^{26}$Mg/$^{24}$Mg (see Figure \ref{fig:chi}).  We took $\Delta\chi^2 = 1$ to be the 
1$\sigma$ confidence limit for determining $^{25}$Mg/$^{24}$Mg or $^{26}$Mg/$^{24}$Mg.  For each region of 
each star, we paired an uncertainty to the optimized value for $^{25}$Mg/$^{24}$Mg or $^{26}$Mg/$^{24}$Mg
and a weighted mean was calculated giving a single value of $^{24}$Mg:$^{25}$Mg:$^{26}$Mg for each star 
(see Table \ref{tab:param}).  As stated in previous studies of the Mg isotope ratios, we find that the 
ratio $^{25}$Mg/$^{24}$Mg is less certain than $^{26}$Mg/$^{24}$Mg since $^{26}$MgH is less blended with
the strong $^{24}$MgH line.  We also find that isotope ratios from Regions 2 and 3 are less accurate than
from Region 1, as noted by \citet{6752}.  As previously shown by \citet{ml88} and \citet{6752}, Region 1 
tends to give higher ratios than Regions 2 and 3.  In calculating the mean isotope ratio for a given star,
we also determine formal statistical errors ($\Delta\chi^2 = 1$).  However, these errors are very 
small, and neglect
systematic errors from continuum fitting, microturbulence, macroturbulence, identified and unidentified
blends.  Examination of Figures \ref{fig:region1} and \ref{fig:region23} show that it is difficult to
discern by eye differences in the syntheses at or below the level b $\pm$ 3 or c $\pm$ 3 when expressing 
the ratio as $^{24}$Mg:$^{25}$Mg:$^{26}$Mg=(100$-$b$-$c):b:c.  The derived isotope ratios are insensitive
to the adopted stellar parameters (see \citealt{mghdwarf} for a discussion of uncertainties).  For 
HIP 18946 we adopted a microturbulence of 0.8 km s$^{-1}$ (originally 0.3 km s$^{-1}$) and measured
the ratio 75:12:13 (originally 76:12:12).  Even for the coolest star HIP 19082 for which the
microturbulence would have the greatest effect, the ratio was 78:11:11 (originally 80:10:10) when
the microturbulence was changed to 0.8 km s$^{-1}$ (from 0.3 km s$^{-1}$).

\begin{figure}
\epsscale{1.0}
\plotone{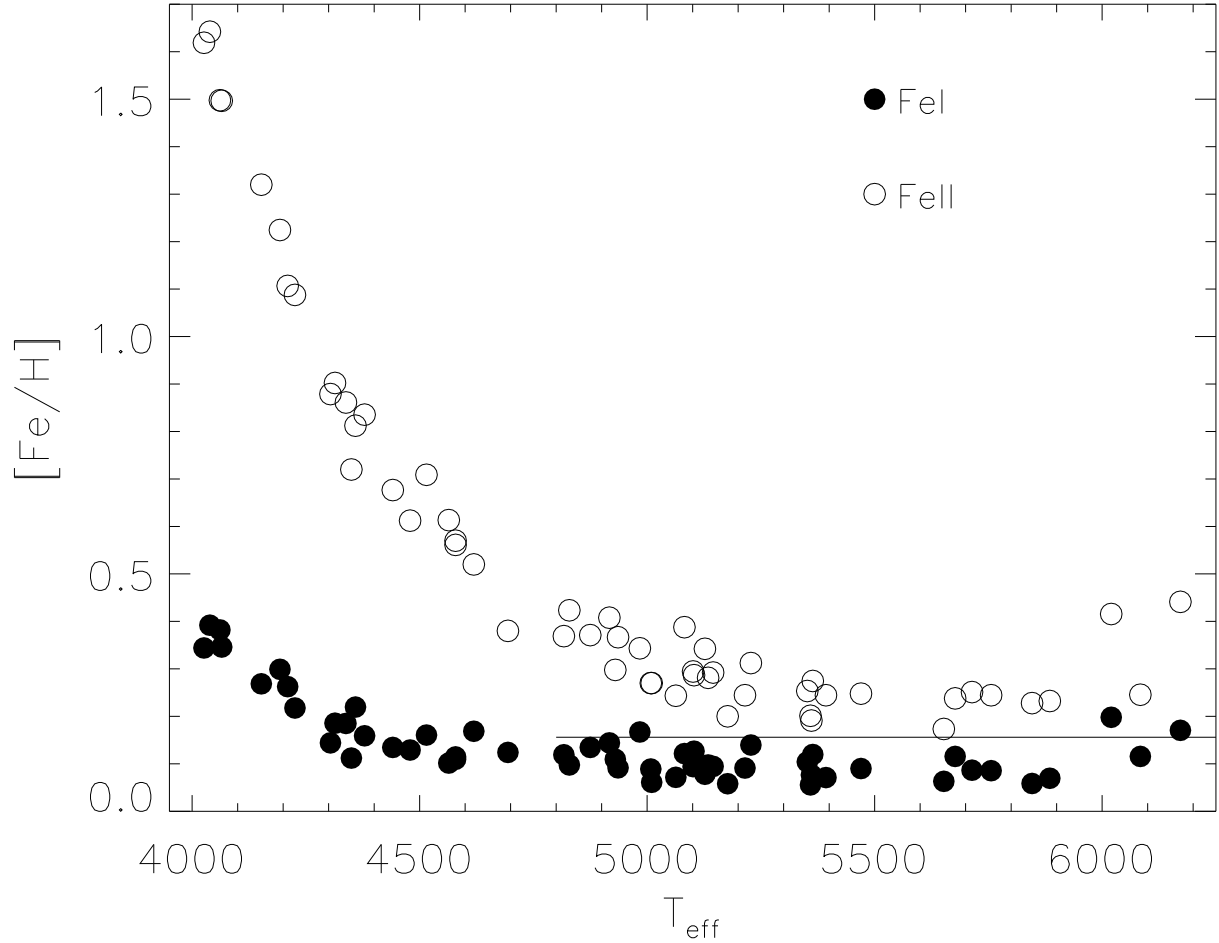}
\caption{Mean Fe abundance from Fe\,{\sc i} and Fe\,{\sc ii} lines versus \teff.  
Above 4700K, the difference between Fe\,{\sc i} and Fe\,{\sc ii} is constant at about
0.2 dex while below 4700K, the difference between Fe\,{\sc i} and Fe\,{\sc ii} 
increases with decreasing metallicity.
The solid line is the mean value of Fe 
based only on stars with \teff$>4700$K.
\label{fig:fe.all}}
\end{figure}

\begin{figure}
\epsscale{1.0}
\plotone{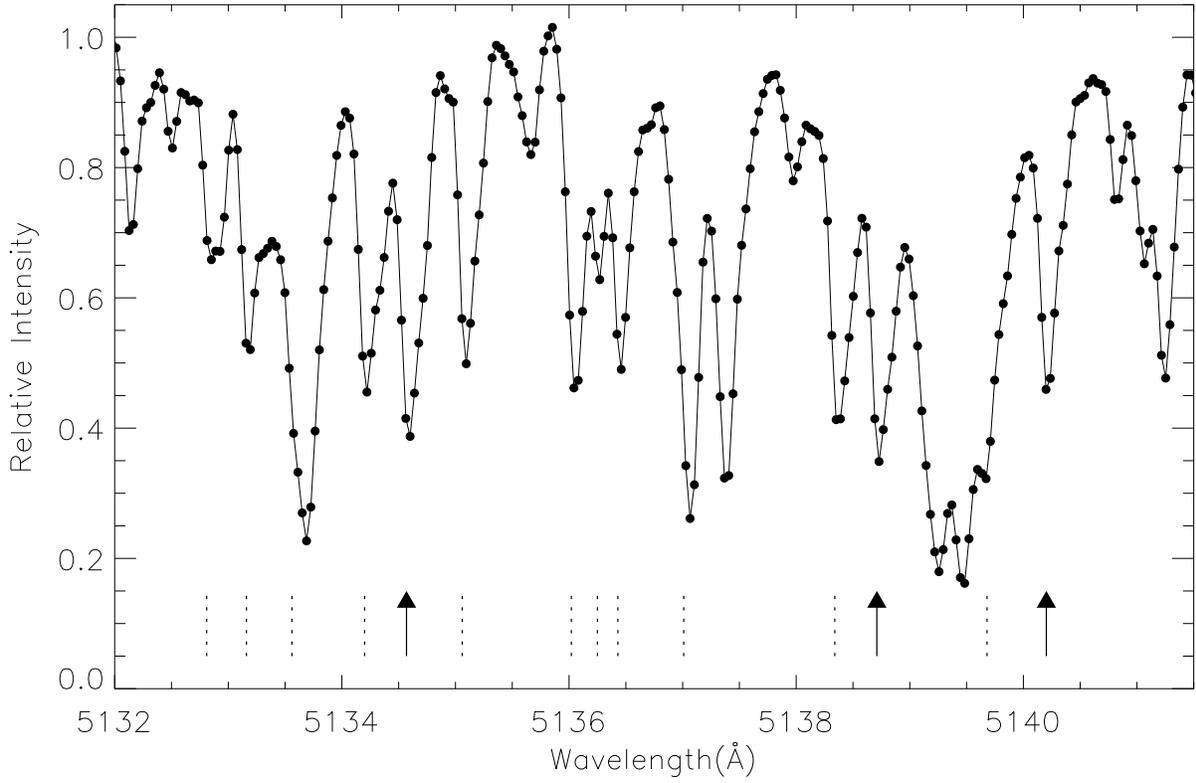}
\caption{Spectrum of HIP 18322 showing the positions of
various MgH lines.  While most of the MgH lines are unsuitable
for isotopic analysis, 3 features we use to derive the isotope
ratios are marked by arrows.
\label{fig:mgh.all}}
\end{figure}

\begin{figure}
\epsscale{1.0}
\plotone{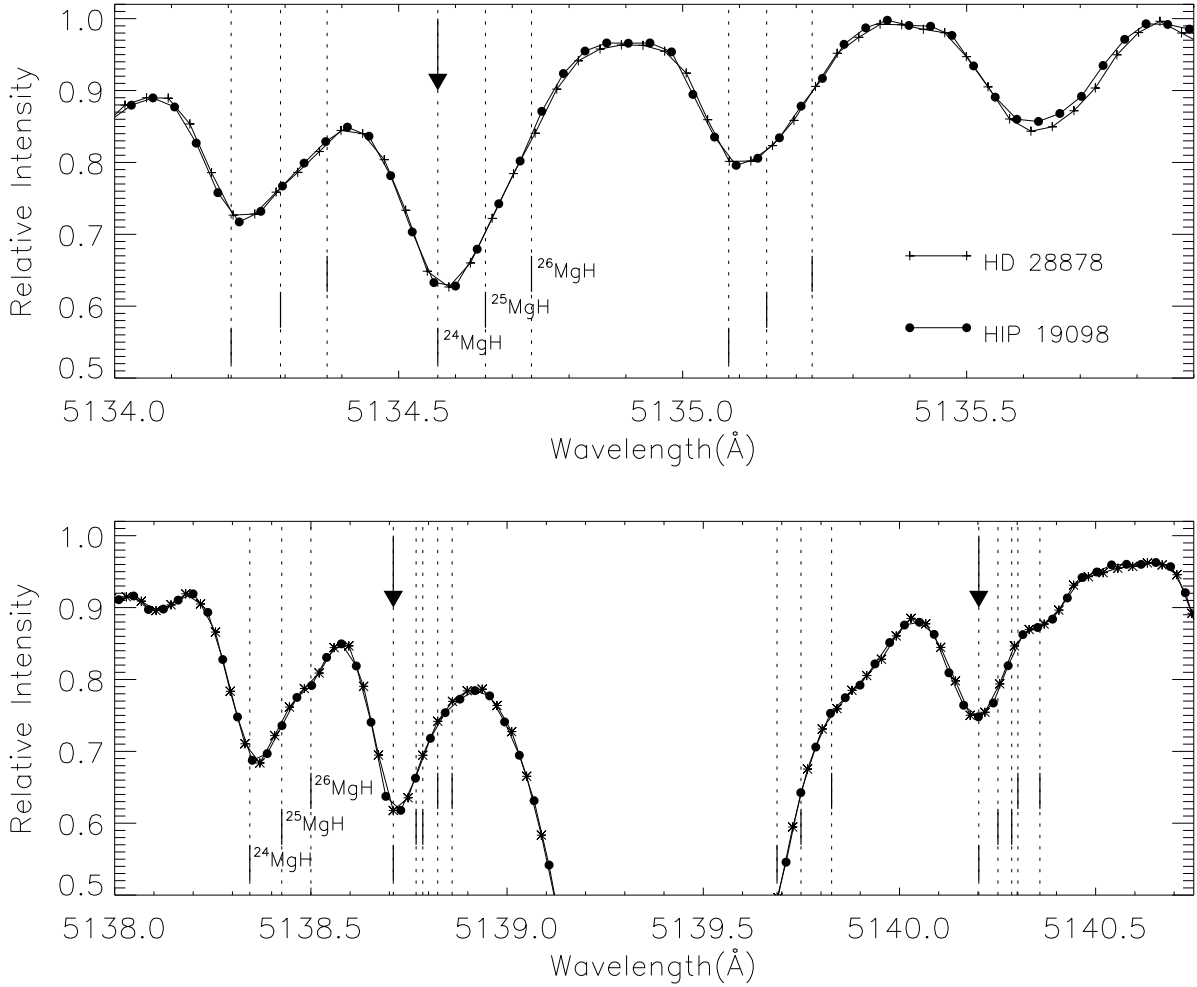}
\caption{Spectra of HD 28878 and HIP 19098.  The positions of 
the $^{24}$MgH, $^{25}$MgH, and $^{26}$MgH lines are shown and the
lines used to derive the isotope ratios are highlighted by
arrows.  There are no discernable differences in the profiles
of the MgH lines suggesting that both stars have similar
Mg isotope ratios.
\label{fig:mgh.comp}}
\end{figure}

\begin{figure}
\epsscale{1.0}
\plotone{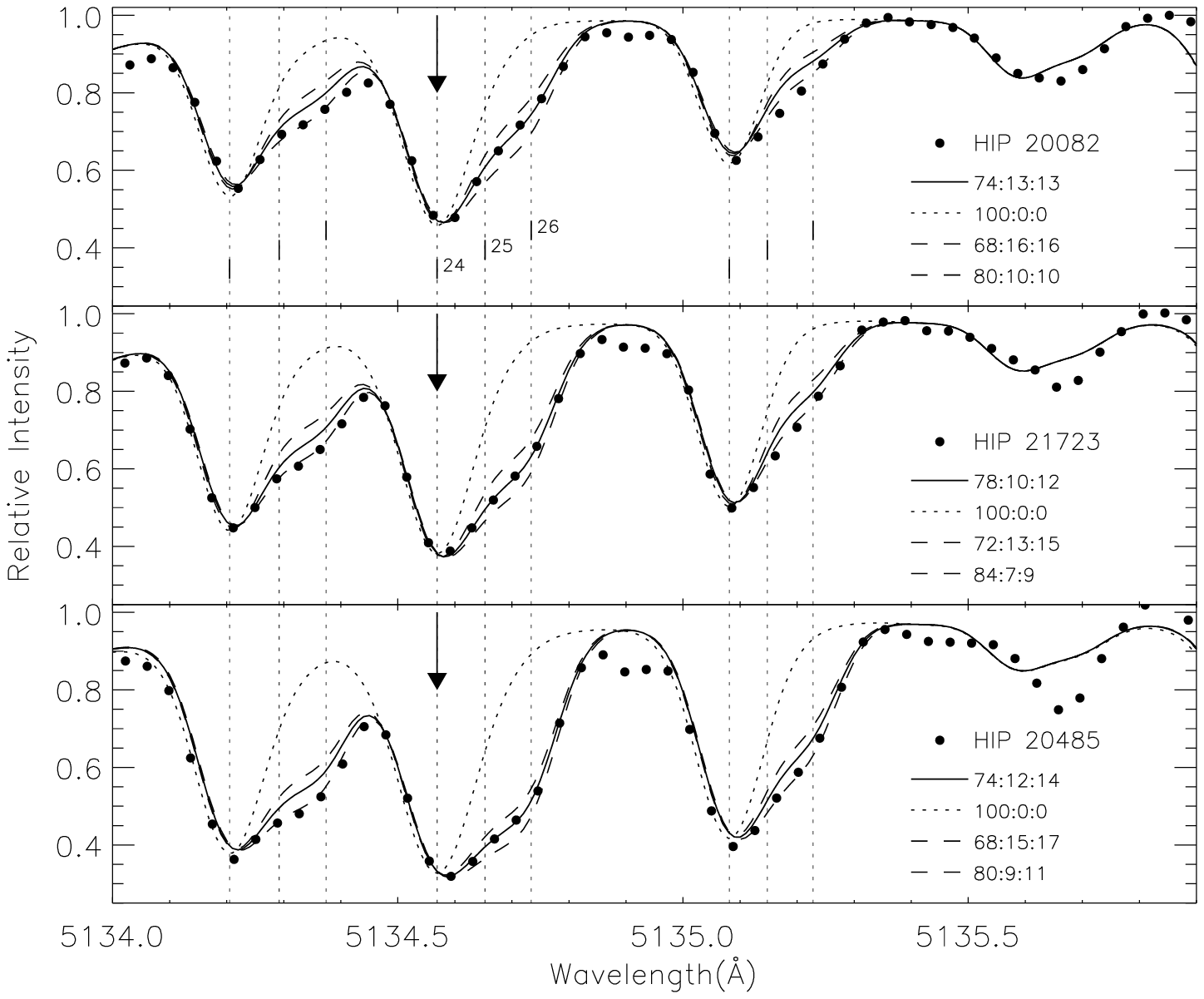}
\caption{Spectra of HIP 20082, HIP 21723, and HIP 20485.  
The feature we are fitting is highlighted by the arrow.  The
positions of the $^{24}$MgH, $^{25}$MgH, and $^{26}$MgH lines
are indicated by dashed lines.  The closed circles represent
the observed spectra, the best fit is the solid line, and 
unsatisfactory ratios are also shown.
\label{fig:region1}}
\end{figure}

\begin{figure}
\epsscale{1.0}
\plotone{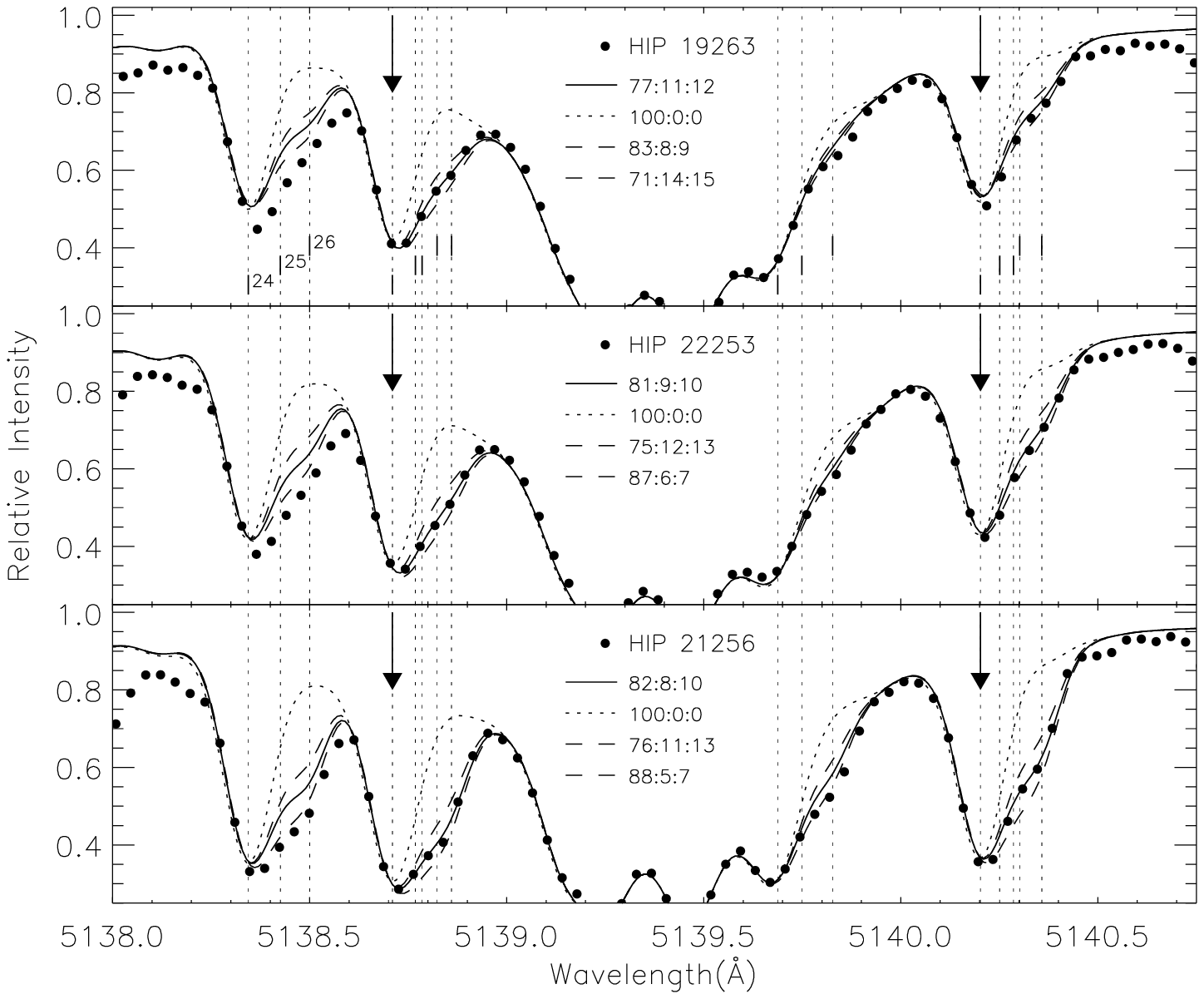}
\caption{Spectra of HIP 19263, HIP 22253, and HIP 21256.  
The features we are fitting are highlighted by the arrows.  The
positions of the $^{24}$MgH, $^{25}$MgH, and $^{26}$MgH lines
are indicated by dashed lines.  The closed circles represent
the observed spectra, the best fit is the solid line, and 
unsatisfactory ratios are also shown.
\label{fig:region23}}
\end{figure}

\begin{figure}
\epsscale{1.0}
\plotone{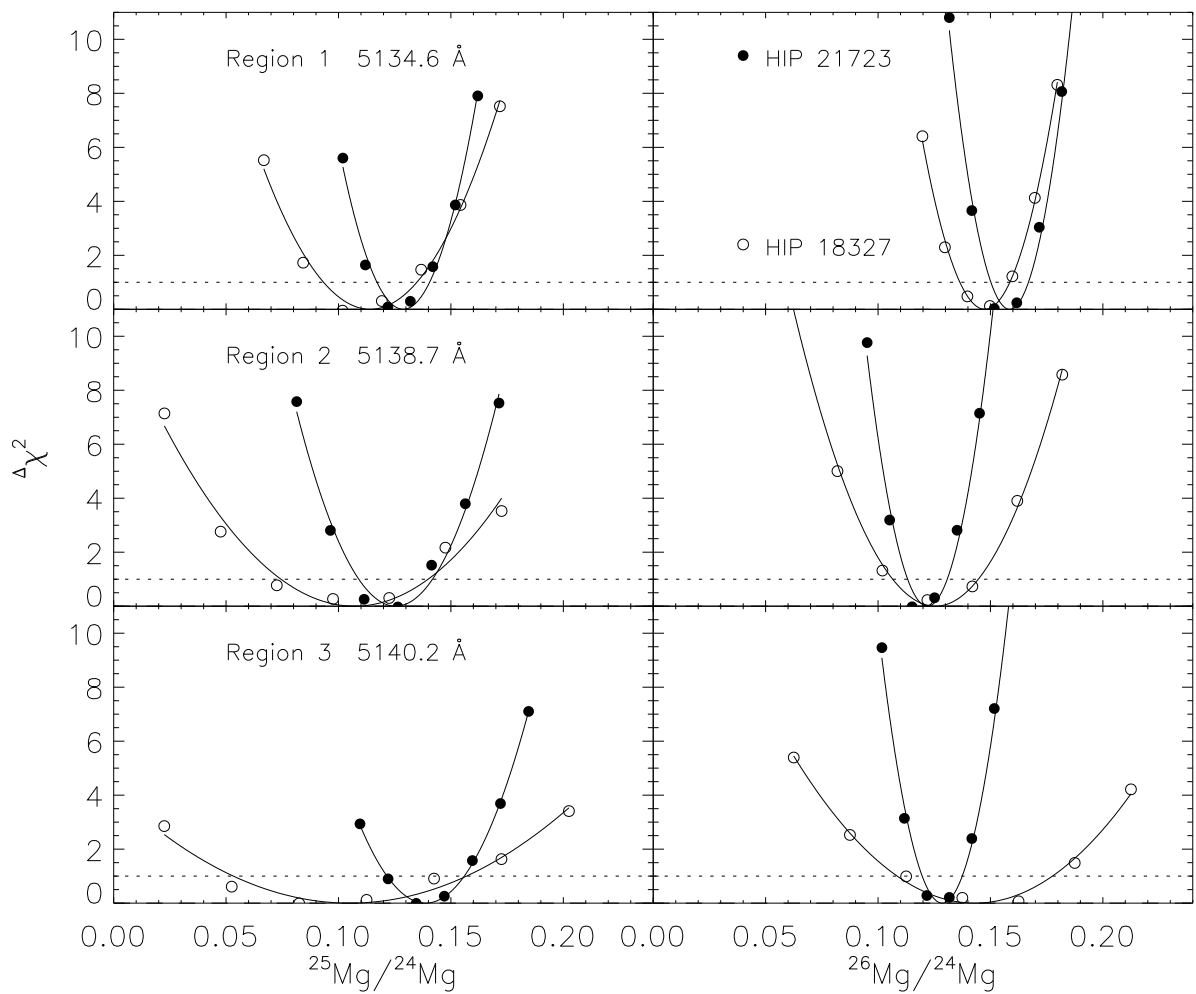}
\caption{Variation in $\Delta\chi^2$ fit for
HIP 21723 and HIP 18327 for $^{25}$Mg/$^{24}$Mg
(left panels) and $^{26}$Mg/$^{24}$Mg (right panels).
The upper, middle, and lower panels show the
$\chi^2$ variation for Region 1 (5134.6\AA), 
Region 2 (5138.7\AA), and Region 3 (5140.2\AA).  
The line indicating 1$\sigma$ errors ($\Delta \chi^2 = 1$) 
is shown.
\label{fig:chi}}
\end{figure}

\begin{figure}
\epsscale{1.0}
\plotone{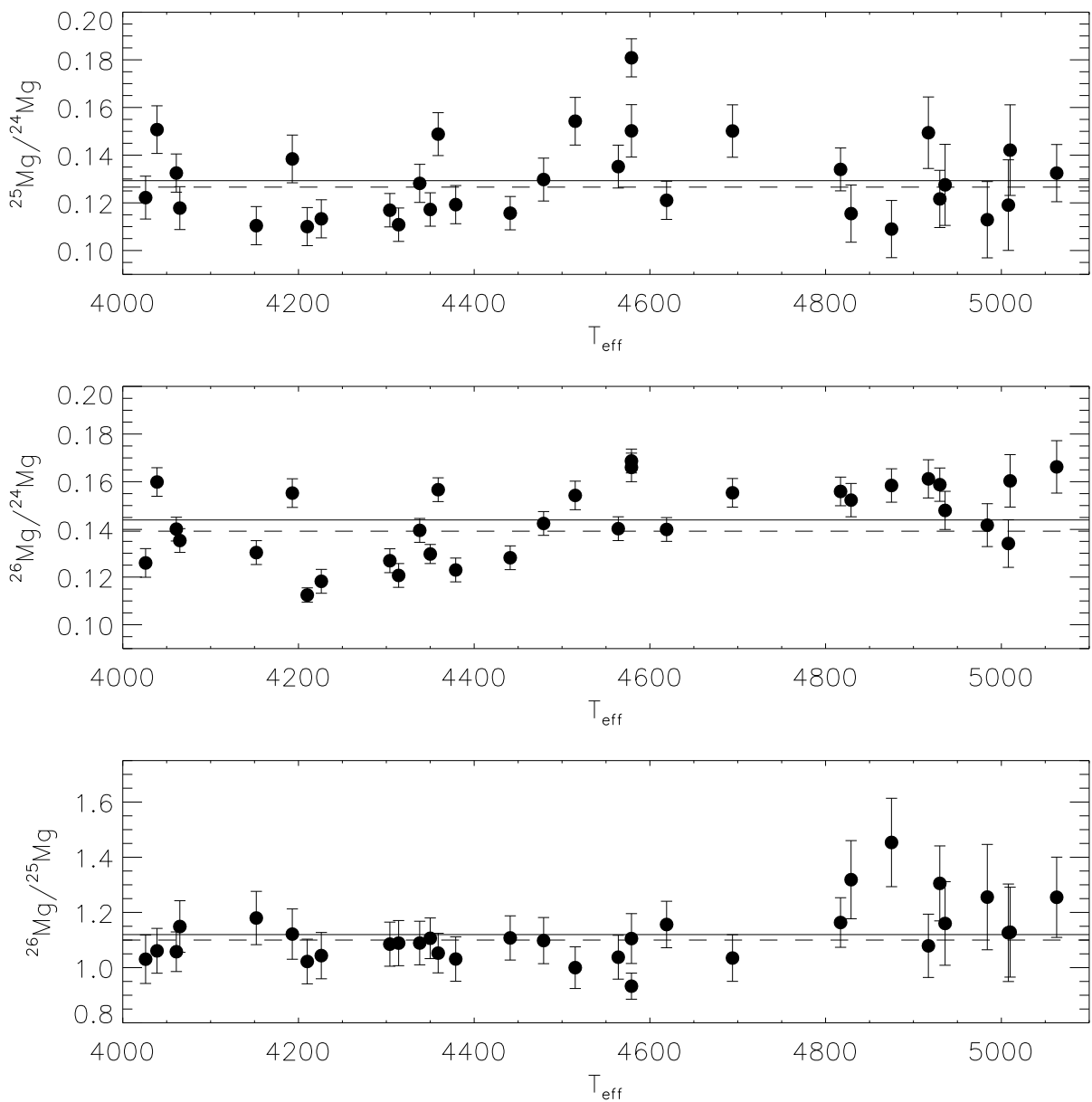}
\caption{Mg isotope ratio $^{25}$Mg/$^{24}$Mg (upper), 
$^{26}$Mg/$^{24}$Mg (middle), and $^{26}$Mg/$^{25}$Mg (lower) 
versus \teff.  The solid line
represents the mean value and the dashed line represents
the solar value.  The error bars show the formal statistical
errors which almost certainly underestimate the true errors
(see discussion in text).  There is no significant trend of any
isotope ratio with \teff.
\label{fig:mg.teff}}
\end{figure}

In Figure \ref{fig:mg.teff}, we plot the Mg isotope ratios $^{25}$Mg/$^{24}$Mg, $^{26}$Mg/$^{24}$Mg, and
$^{26}$Mg/$^{25}$Mg versus \teff.  None of the isotope ratios show a significant trend with \teff.  
The mean ratio for the Hyades is $^{24}$Mg:$^{25}$Mg:$^{26}$Mg = 78.6:10.1:11.3, almost identical to
the solar ratio 78.99:10.00:11.01 \citep{mghsolar}. This demonstration that
the Hyades and solar system isotopic ratios are in good
agreement is not a surprise given the small difference in
composition between the Hyades and Sun.  In usual parlance
[Fe/H] is small (see above) and
[X/Fe] is zero to within small measurement errors for all
elements (except Li and Be) examined.  What may be a surprise given
the results in Figure \ref{fig:fe.all} is that the Hyades Mg isotope ratios are quite
independent of \teff. This result will serve to test
explanations of Figure \ref{fig:fe.all}.

Classical atmospheres assume homogenous layers.  Should the real atmosphere consist of hot and cool 
columns, MgH lines will be strong in the cool columns and weak in the hot columns.  The continuum from hot
columns will weaken the MgH lines from the cool columns.  The analysis of the combined spectrum 
using a classical atmosphere will lead to an underestimate of the saturation and an overestimate 
of the ratios $^{25}$Mg/$^{24}$Mg and $^{26}$Mg/$^{24}$Mg.  \citet{lambert71} showed that this 
effect was responsible for producing artificially high Mg isotope ratios reported from sunspot 
spectra.  Thus, it would be useful to measure the Mg isotope ratios using a three-dimensional 
hydrodynamic model atmosphere.  
The youth of the Hyades may give rise to a significant fraction of starspot coverage.  

\begin{figure}
\epsscale{1.0}
\plotone{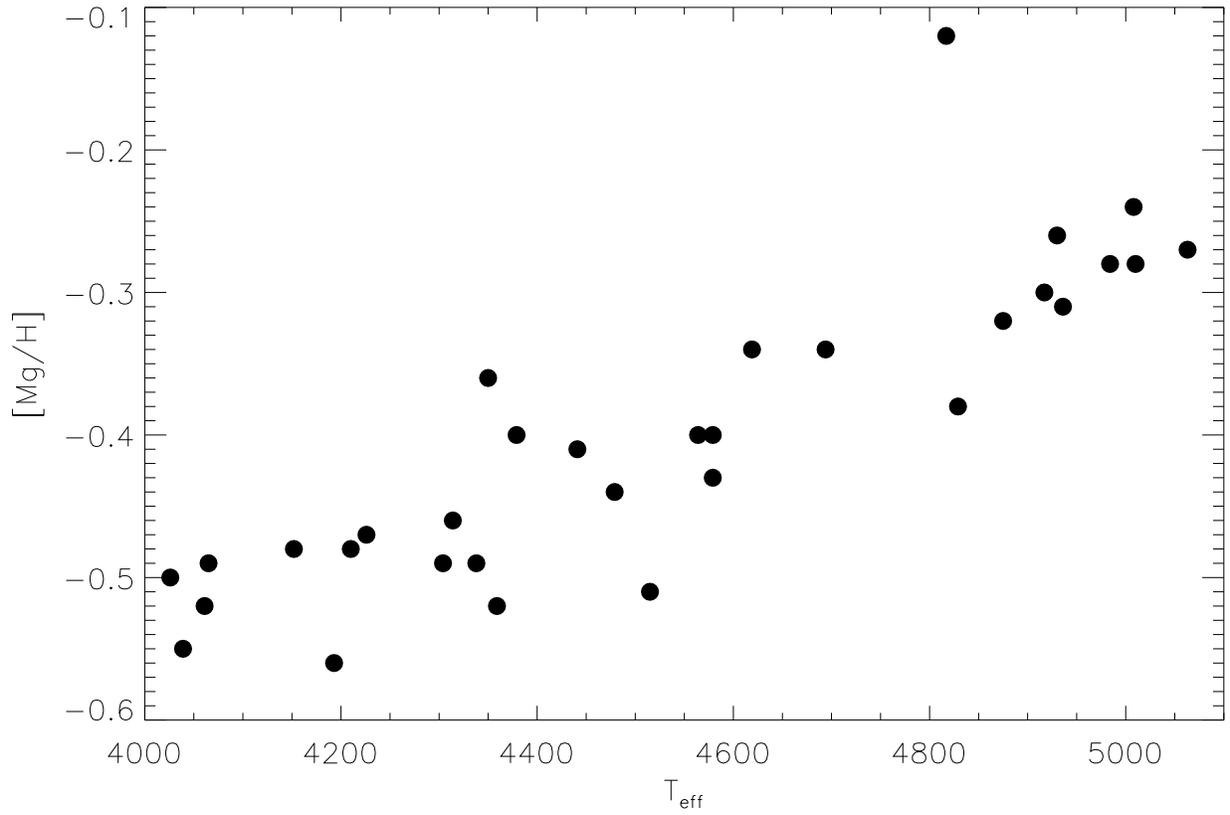}
\caption{Mg abundance (derived from MgH lines) versus \teff.
There is a significant trend of [Mg/H] with \teff~as well
as an offset with respect to the \citet{paulson03} values.  
(Paulson et al.\ derive [Mg/Fe]$\simeq$0.0 which corresponds
to [Mg/H]$\simeq$0.16.)  
\label{fig:elmg.teff}}
\end{figure}

Normally we avoid using the MgH features to determine the Mg abundance.  Slight variations in the 
adopted temperature can result in large changes in the derived Mg abundance.  Uncertainties in the 
absolute $gf$-values of the MgH lines and the molecule's dissociation energy will also introduce a 
systematic offset in the derived abundances.  
We plot the Mg abundances versus \teff~in Figure \ref{fig:elmg.teff} and find that the abundance 
decreases with decreasing \teff.  If we plot [Mg/Fe\,{\sc i}] (where Fe\,{\sc i} 
is the iron abundance from neutral species) the trend of Mg with \teff~would be even more
pronounced.  For HIP 19098 
(\teff=4978K) and HIP 18946 (\teff=4485K), we increased the adopted \teff~by 150K and found that the
Mg abundance increased by about 0.3 dex.  Therefore, in the warmer stars, the 150K 
adjustment brings the Mg abundance from MgH lines into agreement with the \citeauthor{paulson03} 
values as well as reproducing ionization equilibrium.  However, in our cooler stars, 
the 150K adjustment does not result in agreement with the \citeauthor{paulson03} Mg abundances, the
trend of [Mg/H] with \teff~does not disappear, and iron ionization equilibrium is not satisfied.  
(The \citeauthor{paulson03} Mg abundances are derived from atomic Mg lines.)
The anticorrelation seen in Figures \ref{fig:fe.all} and \ref{fig:elmg.teff} suggests that
the mechanisms responsible for the apparent overionization of Fe also affect the 
Mg abundances in the coolest stars.

\section{Concluding remarks}

We measured the Mg isotope ratios in 32 dwarfs with 4000K $\le$ \teff~$\le$ 5000K.  Our goal was to
investigate whether our analysis was susceptible to problems in the model atmospheres or analysis.
In \citet{mghdwarf} the two stars with high ratios of $^{25}$Mg/$^{24}$Mg were rather cool and there was a 
hint that the Mg isotope ratios showed a trend with \teff.  (We note that these two stars were metal-poor 
relative to the Hyades.)  Use of the Mg isotope ratios to investigate the 
chemical history of our Galaxy would be severely weakened if a temperature bias exists.  It is reassuring 
that in the Hyades stars, we find no trend with \teff.  The stars in this study have MgH
lines of comparable strength to the two cool stars with high isotope ratios identified by \citet{mghdwarf}.  
The mean Mg isotope ratio for the
Hyades is in excellent agreement with the solar value which is unsurprising since the Hyades have been
shown to have elemental abundance ratios in accord with the solar ratios.  

We also determined the stellar parameters and iron abundances for 56 Hyades dwarfs with 4000K $\le$ 
\teff~$\le$ 6200K.  For stars warmer than 4700K, we found a mean Fe abundance in good agreement with
previous studies.  The Fe abundance from Fe\,{\sc ii} lines was slightly higher than from Fe\,{\sc i} lines.
Modest adjustments to the stellar parameters would result in the same Fe abundance from neutral and
ionized lines.  Below 4700K, we found overionization of Fe where the effect increased with decreasing \teff.
Unrealistic changes to the adopted stellar parameters would be required to satisfy ionization equilibrium.
Similar problems have been seen by \citet{feltzing98}, \citet{schuler03}, and 
Allende Prieto et.\ al (2003, private communication) and all studies rely upon classical model atmospheres.  
None of the previous studies considered dwarfs as cool as those investigated in this study.  We identify 
the problem of apparent overionization in cool Hyades dwarfs but
offer no definitive explanation.  It is likely that inadequacies in the model atmospheres and neglect of 
Non-LTE and/or other effects are to blame.  Even though the models fail to satisfy ionization equilibrium, 
our Mg isotope ratios do not show any such problem.  Abundance ratios determined via the analysis of 
identical transitions in essentially
identical species are insensitive to errors in the model atmospheres or analysis.  

In order to further investigate this problem of overionization of Fe, we intend to undertake a more complete
abundance analysis of the Hyades focusing upon more elements (neutral and singly-charged ions where possible)
as well as abundances from molecular lines.  A parallel study of members of the Pleiades, other open clusters,
and field stars should also be conducted.  Such a study may reveal how age and metallicity influence the
mechanism responsible for the overionization of Fe.

\acknowledgments

We thank Bill Cochran and Artie Hatzes for providing the data.  
We thank the anonymous referee for helpful comments that improved the
clarity of the paper.
We acknowledge support from the Robert A. Welch
Foundation of Houston, Texas.  
This research has made use of the SIMBAD database,
operated at CDS, Strasbourg, France and
NASA's Astrophysics Data System.

\clearpage

\clearpage

\clearpage

\clearpage

\clearpage

\clearpage

\clearpage

\clearpage

\clearpage

\clearpage

\clearpage

\clearpage

\begin{deluxetable}{lccc} 
\tabletypesize{\scriptsize}
\tablecolumns{4} 
\tablewidth{0pc} 
\tablecaption{Atomic line list \label{tab:lines}}
\tablehead{ 
\colhead{Species}          &
\colhead{Wavelength (\AA)} &
\colhead{$\chi$ (eV)}        &
\colhead{log $gf$}
}

\startdata 
Fe I & 5322.041 & 2.28 & $-$2.84 \\
Fe I & 5811.919 & 4.14 & $-$2.43 \\
Fe I & 5853.161 & 1.49 & $-$5.28 \\
Fe I & 5855.086 & 4.61 & $-$1.60 \\
Fe I & 5856.096 & 4.30 & $-$1.64 \\
Fe I & 5858.785 & 4.22 & $-$2.26 \\
Fe I & 5927.797 & 4.65 & $-$1.09 \\
Fe I & 5933.803 & 4.64 & $-$2.23 \\
Fe I & 5940.997 & 4.18 & $-$2.15 \\
Fe I & 5956.706 & 0.86 & $-$4.61 \\
Fe I & 5969.578 & 4.28 & $-$2.73 \\
Fe I & 6019.364 & 3.57 & $-$3.36 \\
Fe I & 6027.051 & 4.08 & $-$1.09 \\
Fe I & 6054.080 & 4.37 & $-$2.31 \\
Fe I & 6105.130 & 4.55 & $-$2.05 \\
Fe I & 6151.618 & 2.18 & $-$3.29 \\
Fe I & 6157.728 & 4.08 & $-$1.11 \\
Fe I & 6159.380 & 4.61 & $-$1.97 \\
Fe I & 6165.360 & 4.14 & $-$1.47 \\
Fe I & 6173.336 & 2.22 & $-$2.88 \\
Fe II & 4491.407 & 2.86 & $-$2.49 \\
Fe II & 4508.290 & 2.86 & $-$2.31 \\
Fe II & 4620.520 & 2.83 & $-$3.23 \\
Fe II & 5197.559 & 3.23 & $-$2.25 \\
Fe II & 5264.810 & 3.23 & $-$3.15 \\
Fe II & 5325.559 & 3.22 & $-$3.17 \\
Fe II & 5414.046 & 3.22 & $-$3.62 \\
Fe II & 5425.247 & 3.20 & $-$3.21 \\
Fe II & 6149.246 & 3.89 & $-$2.72 \\
\enddata 

\end{deluxetable} 

\begin{deluxetable}{lccccccccc} 
\tabletypesize{\scriptsize}
\tablecolumns{10} 
\tablewidth{0pc} 
\tablecaption{The program stars  \label{tab:param}}
\tablehead{ 
\colhead{Name}          &
\colhead{$B-V$\tablenotemark{a}}           &
\colhead{\teff}         &
\colhead{log g}         &
\colhead{$\xi_t$}       &
\colhead{Macro}         &
\colhead{Region1}       &
\colhead{Region2}       &
\colhead{Region3}       &
\colhead{Final ratio\tablenotemark{b}}    \\
\colhead{}              &
\colhead{}              &
\colhead{(K)}           &
\colhead{(cm s$^{2}$)}  &
\multicolumn{2}{c}{km s$^{-1}$} &
\multicolumn{4}{c}{$^{24}$Mg:$^{25}$Mg:$^{26}$Mg} 
}
\startdata 
HIP 20557 & 0.52 & 6172 & 4.42 & 0.7 & \nodata & \nodata & \nodata & \nodata & \nodata \\
HIP 21112 & 0.54 & 6084 & 4.44 & 0.6 & \nodata & \nodata & \nodata & \nodata & \nodata \\
HIP 20237 & 0.56 & 6020 & 4.46 & 0.6 & \nodata & \nodata & \nodata & \nodata & \nodata \\
HIP 19148 & 0.59 & 5885 & 4.49 & 0.6 & \nodata & \nodata & \nodata & \nodata & \nodata \\
HIP 20577 & 0.60 & 5846 & 4.49 & 0.6 & \nodata & \nodata & \nodata & \nodata & \nodata \\
HIP 21317 & 0.63 & 5756 & 4.51 & 0.6 & \nodata & \nodata & \nodata & \nodata & \nodata \\
HIP 19786 & 0.64 & 5714 & 4.52 & 0.7 & \nodata & \nodata & \nodata & \nodata & \nodata \\
HIP 20741 & 0.66 & 5652 & 4.53 & 0.7 & \nodata & \nodata & \nodata & \nodata & \nodata \\
HIP 19793 & 0.66 & 5677 & 4.52 & 0.7 & \nodata & \nodata & \nodata & \nodata & \nodata \\
HIP 20146 & 0.72 & 5470 & 4.56 & 0.7 & \nodata & \nodata & \nodata & \nodata & \nodata \\
HIP 20130 & 0.75 & 5393 & 4.57 & 0.5 & \nodata & \nodata & \nodata & \nodata & \nodata \\
HIP 20480 & 0.76 & 5359 & 4.57 & 0.6 & \nodata & \nodata & \nodata & \nodata & \nodata \\
HIP 23498 & 0.77 & 5352 & 4.58 & 0.5 & \nodata & \nodata & \nodata & \nodata & \nodata \\
HIP 24923 & 0.77 & 5334 & 4.57 & 0.5 & \nodata & \nodata & \nodata & \nodata & \nodata \\
HIP 20949 & 0.77 & 5331 & 4.57 & 0.6 & \nodata & \nodata & \nodata & \nodata & \nodata \\
HIP 21741 & 0.81 & 5198 & 4.59 & 0.5 & \nodata & \nodata & \nodata & \nodata & \nodata \\
HIP 19934 & 0.81 & 5215 & 4.60 & 0.3 & \nodata & \nodata & \nodata & \nodata & \nodata \\
HIP 20951 & 0.83 & 5177 & 4.60 & 0.6 & \nodata & \nodata & \nodata & \nodata & \nodata \\
HIP 22380 & 0.83 & 5145 & 4.61 & 0.6 & \nodata & \nodata & \nodata & \nodata & \nodata \\
HIP 20850 & 0.84 & 5127 & 4.61 & 0.4 & \nodata & \nodata & \nodata & \nodata & \nodata \\
HIP 16529 & 0.84 & 5104 & 4.61 & 0.4 & \nodata & \nodata & \nodata & \nodata & \nodata \\
HIP 13806 & 0.86 & 5073 & 4.61 & 0.6 & \nodata & \nodata & \nodata & \nodata & \nodata \\
HIP 20492 & 0.86 & 5101 & 4.61 & 0.6 & \nodata & \nodata & \nodata & \nodata & \nodata \\
HIP 20978 & 0.87 & 5082 & 4.62 & 0.4 & \nodata & \nodata & \nodata & \nodata & \nodata \\
HD 29159 & 0.87 & 5063 & 4.62 & 0.6 & 3.25 & 77:10:13 & 75:13:12 & 80:08:12 & 77:10:13 \\
HD 28878 & 0.89 & 5010 & 4.63 & 0.6 & 3.75 & 70:16:14 & 78:11:11 & 77:09:13 & 74:13:13 \\
HIP 19098 & 0.89 & 4978 & 4.63 & 0.4 & 4.50 & 77:12:11 & 81:10:09 & 81:05:14 & 79:10:11 \\
HIP 18327 & 0.90 & 4984 & 4.63 & 0.3 & 4.25 & 79:09:12 & 81:09:10 & 80:08:12 & 80:09:11 \\
HIP 16908 & 0.92 & 4900 & 4.64 & 0.3 & 1.50 & 76:11:13 & 78:10:12 & 81:07:12 & 78:10:12 \\
HD 28977 & 0.92 & 4936 & 4.64 & 0.5 & 3.75 & 77:12:11 & 79:10:11 & 81:06:13 & 78:10:12 \\
HIP 13976 & 0.93 & 4875 & 4.65 & 0.4 & 2.25 & 76:12:12 & 82:05:13 & 80:08:12 & 78:09:13 \\
HIP 20827 & 0.93 & 4917 & 4.64 & 0.4 & 3.25 & 74:13:13 & 78:10:12 & 78:10:12 & 77:11:12 \\
HIP 23312 & 0.96 & 4799 & 4.66 & 0.3 & 2.25 & 78:10:12 & 80:09:11 & 79:09:14 & 79:09:12 \\
HIP 20082 & 0.98 & 4817 & 4.66 & 0.4 & 1.00 & 74:13:13 & 78:09:13 & 80:09:11 & 78:10:12 \\
HIP 19263 & 1.01 & 4694 & 4.68 & 0.3 & 2.00 & 75:12:13 & 77:11:12 & 79:11:10 & 76:12:12 \\
HIP 20563 & 1.05 & 4619 & 4.69 & 0.3 & 1.25 & 76:11:13 & 82:08:10 & 80:10:10 & 79:10:11 \\
HIP 18322 & 1.07 & 4540 & 4.69 & 0.3 & 1.50 & 71:15:14 & 78:11:11 & 77:12:11 & 74:13:13 \\
HIP 22654 & 1.07 & 4540 & 4.69 & 0.3 & 1.75 & 75:11:14 & 77:10:13 & 76:13:11 & 76:11:13 \\
HIP 21723 & 1.07 & 4534 & 4.70 & 0.3 & 1.25 & 78:10:12 & 80:10:10 & 79:11:10 & 78:11:11 \\
HIP 18946 & 1.10 & 4485 & 4.70 & 0.3 & 1.25 & 75:12:13 & 79:10:11 & 76:13:11 & 76:12:12 \\
HIP 22253 & 1.11 & 4449 & 4.71 & 0.3 & 1.75 & 77:10:13 & 81:09:10 & 79:11:10 & 79:10:11 \\
HIP 15563 & 1.13 & 4411 & 4.71 & 0.3 & 1.50 & 75:12:13 & 83:08:09 & 83:08:09 & 81:09:10 \\
HIP 20762 & 1.15 & 4359 & 4.73 & 0.3 & 1.25 & 74:12:14 & 78:11:11 & 77:12:11 & 77:11:12 \\
HIP 18018 & 1.16 & 4349 & 4.72 & 0.3 & 1.00 & 77:10:13 & 83:09:08 & 82:09:09 & 80:10:10 \\
HIP 22271 & 1.17 & 4320 & 4.73 & 0.3 & 1.00 & 77:11:12 & 84:07:09 & 81:10:09 & 81:09:10 \\
HIP 19207 & 1.18 & 4308 & 4.73 & 0.3 & 1.00 & 76:11:13 & 81:09:10 & 80:11:09 & 79:10:11 \\
HIP 19441 & 1.19 & 4284 & 4.73 & 0.3 & 1.25 & 78:10:12 & 83:08:09 & 83:09:08 & 81:09:10 \\
HIP 21261 & 1.20 & 4274 & 4.73 & 0.3 & 1.00 & 76:11:13 & 82:09:09 & 83:08:09 & 81:09:10 \\
HIP 19808 & 1.20 & 4210 & 4.75 & 0.3 & 1.00 & 81:09:10 & 83:07:10 & 82:10:08 & 82:09:09 \\
HIP 20485 & 1.23 & 4193 & 4.75 & 0.3 & 1.00 & 74:12:14 & 78:09:13 & 79:12:09 & 77:11:12 \\
HIP 21256 & 1.24 & 4196 & 4.74 & 0.3 & 1.25 & 81:09:10 & 82:08:10 & 81:10:09 & 81:09:10 \\
HIP 22177 & 1.28 & 4122 & 4.75 & 0.3 & 1.00 & 75:10:15 & 80:10:10 & 85:07:08 & 80:09:11 \\
HIP 21138 & 1.28 & 4065 & 4.76 & 0.3 & 1.00 & 80:09:11 & 77:09:14 & 81:10:09 & 80:09:11 \\
HIP 19316 & 1.33 & 4031 & 4.76 & 0.3 & 1.00 & 77:11:12 & 79:09:12 & 80:11:09 & 79:10:11 \\
HIP 17766 & 1.34 & 4009 & 4.77 & 0.3 & 1.25 & 75:11:14 & 77:10:13 & 76:14:10 & 76:12:12 \\
HIP 19082 & 1.35 & 3996 & 4.77 & 0.3 & 1.00 & 78:10:12 & 80:09:11 & 81:10:09 & 80:10:10 \\
\enddata 

\tablenotetext{a}{$B-V$ values taken from \citet{ap99}}
\tablenotetext{b}{Weighted mean of the ratios derived for regions 1, 2, and 3 
weighted by the $\chi^2$ errors}

\end{deluxetable}

\begin{deluxetable}{lcrcccc} 
\tabletypesize{\scriptsize}
\tablecolumns{7} 
\tablewidth{0pc} 
\tablecaption{Comparison with \citet{paulson03} and
\citet{debruijne01} \label{tab:comp}}
\tablehead{ 
\colhead{}          &
\multicolumn{2}{c}{This study}  &
\multicolumn{2}{c}{\citet{paulson03}}  &
\multicolumn{2}{c}{\citet{debruijne01}} \\
\colhead{Star} &
\colhead{\teff} &
\colhead{log g} &
\colhead{\teff} &
\colhead{log g} &
\colhead{\teff} &
\colhead{log g}
}
\startdata 
HIP 20557 & 6172 & 4.42 & 6400 & 4.3 & 6275 & 4.41 \\
HIP 21112 & 6084 & 4.44 & 6250 & 4.3 & 6186 & 4.42 \\
HIP 20237 & 6020 & 4.46 & 6200 & 4.3 & 6107 & 4.44 \\
HIP 19148 & 5885 & 4.49 & 6100 & 4.5 & 5983 & 4.46 \\
HIP 20577 & 5846 & 4.49 & 6050 & 4.4 & 5961 & 4.47 \\
HIP 21317 & 5756 & 4.51 & 5900 & 4.4 & 5844 & 4.49 \\
HIP 19786 & 5714 & 4.52 & 5900 & 4.4 & 5812 & 4.50 \\
HIP 20741 & 5652 & 4.53 & 5800 & 4.4 & 5735 & 4.51 \\
HIP 19793 & 5677 & 4.52 & 5750 & 4.4 & 5757 & 4.51 \\
HIP 20146 & 5470 & 4.56 & 5600 & 4.5 & 5553 & 4.54 \\
HIP 20130 & 5393 & 4.57 & 5550 & 4.5 & 5485 & 4.55 \\
HIP 20480 & 5359 & 4.57 & 5500 & 4.5 & 5449 & 4.55 \\
HIP 23498 & 5352 & 4.58 & 5500 & 4.5 & 5429 & 4.56 \\
HIP 24923 & 5334 & 4.57 & 5500 & 4.5 & 5429 & 4.56 \\
HIP 20949 & 5331 & 4.57 & 5500 & 4.5 & 5426 & 4.56 \\
HIP 21741 & 5198 & 4.59 & 5350 & 4.5 & 5303 & 4.57 \\
HIP 19934 & 5215 & 4.60 & 5350 & 4.5 & 5297 & 4.57 \\
HIP 20951 & 5177 & 4.60 & 5300 & 4.5 & 5254 & 4.58 \\
HIP 22380 & 5145 & 4.61 & 5300 & 4.6 & 5249 & 4.58 \\
HIP 20850 & 5127 & 4.61 & 5350 & 4.6 & 5235 & 4.58 \\
HIP 16529 & 5104 & 4.61 & 5250 & 4.6 & 5223 & 4.58 \\
HIP 13806 & 5073 & 4.61 & 5200 & 4.6 & 5196 & 4.58 \\
HIP 20492 & 5101 & 4.61 & 5200 & 4.6 & 5196 & 4.58 \\
HIP 20978 & 5082 & 4.62 & 5250 & 4.6 & 5172 & 4.58 \\
HD 29159 & 5063 & 4.62 & 5000 & 4.6 & \nodata & \nodata \\
HD 28878 & 5010 & 4.63 & 5150 & 4.6 & \nodata & \nodata \\
HIP 19098 & 4978 & 4.63 & 5150 & 4.6 & 5114 & 4.59 \\
HIP 18327 & 4984 & 4.63 & 5050 & 4.6 & 5103 & 4.59 \\
HIP 16908 & 4900 & 4.64 & 5050 & 4.6 & 5054 & 4.59 \\
HD 28977 & 4936 & 4.64 & 5150 & 4.6 & \nodata & \nodata \\
HIP 13976 & 4875 & 4.65 & 5000 & 4.6 & 5034 & 4.60 \\
HIP 20827 & 4917 & 4.64 & 5050 & 4.6 & 5028 & 4.60 \\
HIP 23312 & 4799 & 4.66 & 5100 & 4.6 & 4970 & 4.60 \\
HIP 20082 & 4817 & 4.66 & 4900 & 4.6 & 4924 & 4.61 \\
HIP 19263 & 4694 & 4.68 & \nodata & \nodata & 4874 & 4.61 \\
HIP 20563 & 4619 & 4.69 & \nodata & \nodata & 4785 & 4.62 \\
HIP 18322 & 4540 & 4.69 & \nodata & \nodata & 4749 & 4.62 \\
HIP 22654 & 4540 & 4.69 & \nodata & \nodata & 4749 & 4.62 \\
HIP 21723 & 4534 & 4.70 & \nodata & \nodata & 4744 & 4.62 \\
HIP 18946 & 4485 & 4.70 & \nodata & \nodata & 4703 & 4.63 \\
HIP 22253 & 4449 & 4.71 & \nodata & \nodata & 4672 & 4.63 \\
HIP 15563 & 4411 & 4.71 & \nodata & \nodata & 4640 & 4.64 \\
HIP 20762 & 4359 & 4.73 & \nodata & \nodata & 4611 & 4.64 \\
HIP 18018 & 4349 & 4.72 & \nodata & \nodata & 4586 & 4.64 \\
HIP 22271 & 4320 & 4.73 & \nodata & \nodata & 4560 & 4.65 \\
HIP 19207 & 4308 & 4.73 & \nodata & \nodata & 4550 & 4.65 \\
HIP 19441 & 4284 & 4.73 & \nodata & \nodata & 4527 & 4.65 \\
HIP 21261 & 4274 & 4.73 & \nodata & \nodata & 4515 & 4.65 \\
HIP 19808 & 4210 & 4.75 & \nodata & \nodata & 4500 & 4.65 \\
HIP 20485 & 4193 & 4.75 & \nodata & \nodata & 4438 & 4.66 \\
HIP 21256 & 4196 & 4.74 & \nodata & \nodata & 4425 & 4.66 \\
HIP 22177 & 4122 & 4.75 & \nodata & \nodata & 4320 & 4.66 \\
HIP 21138 & 4065 & 4.76 & \nodata & \nodata & 4312 & 4.66 \\
HIP 19316 & 4031 & 4.76 & \nodata & \nodata & 4129 & 4.67 \\
HIP 17766 & 4009 & 4.77 & \nodata & \nodata & 3980 & 4.71 \\
HIP 19082 & 3996 & 4.77 & \nodata & \nodata & 3944 & 4.72 \\
\enddata 

\end{deluxetable}

\begin{deluxetable}{llrr} 
\tabletypesize{\scriptsize}
\tablecolumns{4} 
\tablewidth{0pc} 
\tablecaption{Abundance dependences on model parameters \label{tab:aberr}}
\tablehead{ 
\colhead{Star}            &
\colhead{Model parameter} &
\colhead{log$\epsilon$(Fe\,{\sc i})}   &
\colhead{log$\epsilon$(Fe\,{\sc ii})}
}
\startdata 
HIP 20557 (\teff=6172) & \teff~+ 100K & 0.06 & $-$0.01 \\
 & log g $-$ 0.2 dex & 0.01 & $-$0.04 \\
 & $\xi$ + 0.2 km s$^{-1}$ & $-$0.03 & $-$0.07 \\
HIP 20146 (\teff=5470) & \teff~+ 100K & 0.07 & $-$0.04 \\
 & log g $-$ 0.2 dex & 0.00 & $-$0.06 \\
 & $\xi$ + 0.2 km s$^{-1}$ & $-$0.03 & $-$0.06 \\
HIP 19098 (\teff=4978) & \teff~+ 100K & 0.02 & $-$0.08 \\
 & log g $-$ 0.2 dex & $-$0.01 & $-$0.09 \\
 & $\xi$ + 0.2 km s$^{-1}$ & $-$0.02 & $-$0.03 \\
HIP 18946 (\teff=4485) & \teff~+ 100K & $-$0.01 & $-$0.14 \\
 & log g $-$ 0.2 dex & $-$0.02 & $-$0.13 \\
 & $\xi$ + 0.2 km s$^{-1}$ & $-$0.01 & $-$0.02 \\
HIP 19082 (\teff=3996) & \teff~+ 100K & $-$0.08 & $-$0.29 \\
 & log g $-$ 0.2 dex & $-$0.07 & $-$0.21 \\
 & $\xi$ + 0.2 km s$^{-1}$ & $-$0.01 & $-$0.01 \\
\enddata 

\end{deluxetable}

\end{document}